\documentclass[prb,amsmath,amssymb,superscriptaddress,twocolumn]{revtex4}
\usepackage{color}
\usepackage{graphicx}
\usepackage{epsfig}
\usepackage{dcolumn}
\usepackage{bm}

\begin{document}
\title{Theory of double Cooper-pair tunneling and light emission mediated by a resonator}

\author{W. T. Morley}
\affiliation{Centre for the Mathematics and Theoretical Physics of Quantum Non-Equilibrium Systems and School of Physics and Astronomy,
             University of Nottingham, Nottingham NG7 2RD, United Kingdom}

\author{A. Di Marco}
\affiliation{Fachbereich Physik, Universit{\"a}t Konstanz,
             D-78457 Konstanz, Germany}
             \affiliation{Department of Microtechnology and Nanoscience (MC2), Chalmers University of Technology, SE-41298 G\"{o}teborg, Sweden}

\author{M. Mantovani}
\affiliation{Fachbereich Physik, Universit{\"a}t Konstanz,
             D-78457 Konstanz, Germany}
\author{P. Stadler}
\affiliation{Department of Microtechnology and Nanoscience (MC2), Chalmers University of Technology, SE-41298 G\"{o}teborg, Sweden}
\author{W. Belzig}
\affiliation{Fachbereich Physik, Universit{\"a}t Konstanz,  D-78457 Konstanz,
             Germany}
\author{G. Rastelli}
\affiliation{Fachbereich Physik, Universit{\"a}t Konstanz,  D-78457 Konstanz,
             Germany}
\affiliation{Zukunftskolleg, Universit{\"a}t Konstanz, D-78457 Konstanz,
             Germany}

\author{A. D. Armour}
\affiliation{Centre for the Mathematics and Theoretical Physics of Quantum
             Non-Equilibrium Systems and School of Physics and Astronomy,
             University of Nottingham, Nottingham NG7 2RD, United Kingdom}
\date{\today}

\begin{abstract}
Photon emission by tunneling electrons can be encouraged by locating a resonator close to the tunnel junction and applying an appropriate voltage-bias. However, studies of normal metals show that the resonator also affects how the charges flow, facilitating processes in which correlated tunneling of two charges produces one photon. We develop a theory to analyze this kind of behavior in Josephson junctions by deriving an effective Hamiltonian describing processes where two Cooper-pairs generate a single photon. We determine the conditions under which the transport is dominated by incoherent tunneling of two Cooper-pairs, whilst also uncovering a regime of coherent double Cooper-pair tunneling. We show that the system can also display an unusual form of photon-blockade and hence could serve as a single-photon source.
\end{abstract}
\maketitle
\section{Introduction}

The tunneling of electrons in mesoscopic conductors or scanning tunneling microscopy (STM) is often accompanied by the generation of photons. Photon emission at a particular frequency can be enhanced, and its detection facilitated, by coupling to a resonator\,\cite{stm1,stm2,holst,stm3,hofheinz,OBE2}. However, the resonator is not simply passive and it can exert a dramatic influence on the charge dynamics, leading even to a change in the effective charge that tunnels. Recent studies\,\cite{OBE2,OBE3,OBE4,OBE5,OBE6} have shown that the presence of an electromagnetic resonator mediates the correlated tunneling of two electrons through a barrier to generate a photon with an energy larger than either electron could individually have provided, a phenomenon known as overbias emission.

\begin{figure}[t]
\centering
{\includegraphics[width=7.5cm]{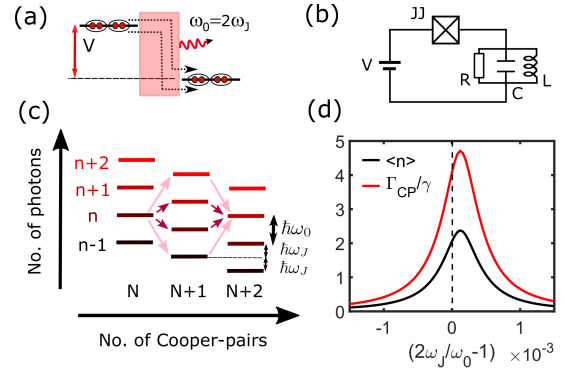}
}
\caption{(a) Cartoon of photon emission accompanying correlated tunneling of two Cooper-pairs through a Josephson junction (JJ). Each Cooper-pair releases energy $\hbar\omega_J=2eV$ so two are required to excite a photon with energy $\hbar\omega_0$ when $\omega_J=\omega_0/2$. (b) Circuit model: A bias voltage $V$ is applied to a JJ in series with a damped $LC$-resonator with frequency $\omega_0$. (c) Level diagram for the number of resonator photons ($n$) and the number of Cooper-pairs passing through the junction ($N$) which illustrates some of the processes which contribute at the resonance.
(d) Time-averaged resonator occupation number, $\langle n\rangle$, and Cooper-pair tunneling rate (scaled by the resonator damping rate), $\Gamma_{CP}/\gamma$, calculated using \eqref{eq:horiginal} and \eqref{eq:me} with parameters $\Delta_0=0.15$, $Q=1500$, $E_J/\hbar\omega_0=0.5$ and $\gamma_{\phi}=0$. The {\emph{ratio}} $\Gamma_{CP}/\gamma\langle n\rangle\simeq 2$, signifying that two Cooper-pairs tunnel for each photon generated as discussed below.
Nonlinearities shift the peak above $\omega_J=\omega_0/2$.}
\label{fig:cpoint}
\end{figure}

In this paper we present a theoretical analysis of overbias emission in superconducting circuits, considering a model circuit consisting of a Josephson junction (JJ) in series with an electrical resonator. When the voltage-bias applied is such that individual tunneling Cooper-pairs provide {\emph{half}} the energy required to generate a photon, charge transport and photon production are dominated by correlated tunneling of {\emph{two}} Cooper-pairs (see Fig.\ \ref{fig:cpoint}). Superconducting circuits are ideally suited to studying higher-order charge tunneling effects. In contrast to a normal conductor, all of the voltage-bias energy of tunneling Cooper-pairs has to be transferred to the electromagnetic environment\,\cite{holst,hofheinz,ingold} and a high-$Q$ resonator can be used to resonantly enhance a wide range of transport processes\,\cite{chen2014}. Furthermore, the photons produced and the charge current flowing are both readily measured\,\cite{hofheinz,chen2014}.

 Whilst photon emission due to tunneling of individual Cooper-pairs has been studied extensively, both experimentally\,\cite{hofheinz,chen2014,cassidy,westig17,amplifier,singlephotonsource,antibunched}  and theoretically\,\cite{paduraiu:12,leppa:13,armour2013,gramich2013,armour2015,simon2015,simon,4thorder,clerk,leppa18,bruder,mendes,hassler} in JJ-resonator systems, higher-order tunneling remains almost completely unexplored. Here we derive an effective Hamiltonian describing tunneling of two Cooper-pairs and use it to investigate the charge transport and photon emission. As the Josephson energy of the junction is increased,  nonlinearity  up-converts the junction Josephson frequency to that of the resonator and the transport evolves from a regime involving tunneling of both one and two Cooper-pairs to one where  incoherent {\emph{double}} Cooper-pair tunneling dominates.  At still larger Josephson energies, the double Cooper-pair tunneling becomes coherent. Although this resonance has been discussed within a classical analysis of the resonator dynamics\,\cite{meister}, a quantum description of the coupled charge-photon dynamics has not been provided until now.

 Our analysis also reveals that double-Cooper pair tunneling leads to a photon-blockade effect\,\cite{photonblockade,sne} which could be exploited as a single-photon source\,\cite{singlephotonsource}. The effect is similar to that seen at single-Cooper pair tunneling resonances\,\cite{gramich2013,clerk}, but the blockade we find occurs at a lower value of the resonator impedance, only slightly higher than that achieved in very recent experiments\,\cite{antibunched}.

\section{Resonator-Junction System}
The model system we consider consists of a LC-resonator with frequency $\omega_0=1/\sqrt{LC}$ in series with a Josephson junction. The resonator could be realised either as the fundamental mode of a superconducting cavity\,\cite{hofheinz,chen2014,cassidy} or as lumped-element oscillator\,\cite{antibunched} (see Fig \ref{fig:cpoint}b). Taking into account the possibility of an additional low-frequency impedance in series with the junction, and assuming the resonator capacitance is much larger than that of the junction\cite{armour2013,meister}, the system can be described by a Hamiltonian of the form\,\cite{gramich2013,bruder}
\begin{equation}
H(t)=\hbar\omega_0\hat{a}^{\dagger}\hat{a}-E_J\cos[\omega_Jt-\varphi+\Delta_0(\hat{a}+\hat{a}^{\dagger})], \label{eq:horiginal}
\end{equation}
where $\hat{a}$ is the lowering operator of the resonator, $\omega_J=2eV/\hbar$ is the Josephson frequency set by the applied voltage, $E_J$ is the Josephson energy and $\Delta_0=(2e^2/\hbar)^{1/2}(L/C)^{1/4}$ gives the zero point flux fluctuations of the resonator in units of the flux quantum. The phase $\varphi$ is conjugate to the number of Cooper pairs, $N$, that have passed through the junction $[\varphi,N]=i$, so that the operator ${\rm{e}}^{ip\varphi}=\sum_N |N+p\rangle\langle N|$ (for integer $p$) describes the transfer of $p$-pairs. The value of $\Delta_0$  is determined by the resonator impedance and although it is much less than unity in standard microwave cavities\,\cite{hofheinz,chen2014,cassidy}, very recent experiments\,\cite{antibunched} utilised a JJ-resonator system with $\Delta_0\simeq 1$.

 We assume that the resonator is subject to losses at a rate $\gamma$ whilst voltage-fluctuations due to the presence of low-frequency impedances in the circuit leads to dephasing of the junction charge at a rate $\gamma_{\varphi}$. In the limit of low temperatures, the master equation is given by\,\cite{gramich2013}
\begin{equation}
\dot{\rho}=-\frac{i}{\hbar}[H,\rho]+\frac{\gamma}{2}{\mathcal{D}}[\hat{a}](\rho)+\frac{\gamma_{\varphi}}{2}{\mathcal{D}}[N](\rho),\label{eq:me}
\end{equation}
where ${\mathcal{D}}[x](\rho)=2x\rho x^{\dagger}-x^{\dagger}x\rho-\rho x^{\dagger}x$. The dephasing term is equivalent to fluctuations in the bias voltage\,\cite{wang2017} and the value of $\gamma_{\varphi}$ is proportional to the zero-frequency voltage noise spectral density. Since typically $\gamma_{\varphi}/\gamma\ll 1$, in many cases the dephasing can be neglected\,\cite{gramich2013,clerk,wang2017,armour2017} and $\varphi$ simply treated as a constant\,\cite{armour2013,wang2017}.

\section{Effective Hamiltonian Description}
We focus on the regime where $2\omega_J\simeq \omega_0$ and processes in which two Cooper-pairs produce a single photon are expected to dominate. Moving to a frame rotating at frequency $2\omega_J$, the corresponding Hamiltonian can be written as
\begin{eqnarray}
\tilde{H}
&=&\hbar\delta \hat{a}^{\dagger}\hat{a}-\frac{\tilde{E}_J}{2}\sum_{q=0}^{\infty}\left[\hat{O}_q{{\rm e}}^{i(2q+1)\omega_Jt} +{\rm{h.c.}}\right],\label{eq:hrf}
\end{eqnarray}
with $\delta=\omega_0-2\omega_J$ and
\begin{eqnarray}
\hat{O}_q&=&:i^q(\hat{a}^{\dagger})^q{\rm{e}}^{-i\varphi}\frac{J_q(2\Delta_0\sqrt{\hat{n}})}{\hat{n}^{q/2}}\nonumber\\
&&+(-i)^{q+1}(\hat{a}^{\dagger})^{q+1}{\rm{e}}^{i\varphi}\frac{J_{q+1}(2\Delta_0\sqrt{\hat{n}})}{\hat{n}^{(q+1)/2}}:,
\end{eqnarray}
where $J_q(z)$ is a Bessel function of order $q$, $\tilde{E}_J=E_J{\rm{e}}^{-\Delta_0^2/2}$, $\hat{n}=\hat{a}^{\dagger}\hat{a}$ is the photon number-operator and $:\dots :$ implies normal ordering.

We obtain an effective (time-independent) Hamiltonian for the system by averaging over short time-scales\,\cite{James} (of order $\sim 1/\omega_J$), making what is in effect a second-order rotating wave approximation\,\cite{RWA2},
\begin{eqnarray}
H_{\rm{eff}}&=&\hbar\delta \hat{a}^{\dagger}\hat{a}+\frac{\tilde{E}^2_J}{4\hbar\omega_J}\sum_{q=0}^{\infty}\frac{\left[\hat{O}_q,\hat{O}^{\dagger}_q\right]}{(2q+1)} \label{eq:5}\\
&=&\left(\hbar\delta +\frac{\tilde{E}_J^2\hat{\mathcal{G}}}{4\hbar\omega_J}\right)\hat{n}-i\frac{\tilde{E}^2_J}{4\hbar\omega_J}\left[\hat{\mathcal{F}}\hat{a}^{\dagger}{\rm{e}}^{2i\varphi}-{\rm{h.c.}}\right], \label{eq:heff}
\end{eqnarray}
where $\hat{\mathcal{F}}(\Delta_0,\hat{n})$ and $\hat{\mathcal{G}}(\Delta_0,\hat{n})$ are higher-order functions of the number operator and $\Delta_0$ (see Appendix \ref{sec:effham} for explicit expressions and a representation in the Fock-state basis). The overall factor of $a^{\dagger}{\rm{e}}^{2i\varphi}$ tells us that the effective Hamiltonian describes coherent processes in which a photon is created in the resonator and {\emph{two}} Cooper-pairs pass through the junction. In terms of the original Hamiltonian, this is a second order-process\,\cite{James} which can be seen as occuring via a range of intermediate (virtual) states as indicated by the sum arising in \eqref{eq:5} (see Fig.\ \ref{fig:cpoint}c).

Equation \ref{eq:heff} also describes a nonlinear shift in the resonator frequency\,\cite{meister} which accounts for the shifted resonance seen in Fig. \ref{fig:cpoint}d. The origin of the frequency shift is rather like the ac-Stark effect, whereby an off-resonant field gives rise to shifts in atomic level spacings without inducing transitions\,\cite{James}. In our case a strong off-resonant drive is present, but it also leads to up conversion (through the nonlinearity) which in turn drives resonant processes. Effective Hamiltonians which are similar in form (though significantly simpler) have been used to describe circuit-QED systems driven by external fields to engineer higher-order photon processes\,\cite{RWA2,six}. In contrast, our effective Hamiltonian describes a higher-order {\emph{charge transport}} process.

Although the full expressions for $\hat{\mathcal{F}}(\Delta_0,\hat{n})$ and $\hat{\mathcal{G}}(\Delta_0,\hat{n})$ are rather cumbersome (see Appendix \ref{sec:effham} for details), if photon numbers are low and $\Delta_0\ll 1$ 
an expansion in which only the lowest-order terms in $\Delta_0$ are retained is sufficient, leading to
\begin{equation}
H^{(0)}_{\rm{eff}}=\hbar\delta'\hat{n}+i\frac{\tilde{E}^2_J\Delta_0^3}{3\hbar\omega_J}\left[\hat{a}{\rm{e}}^{-2i\varphi}-\hat{a}^{\dagger}{\rm{e}}^{2i\varphi}\right], \label{eq:H0}
\end{equation}
with $\delta'=\delta+8\tilde{E}_J^2\Delta_0^4/(15\hbar^2\omega_J)$.

\section{Average photon numbers and charge current}

The photonic properties of the system in the low photon-number regime are readily obtained using (\ref{eq:H0}) and (\ref{eq:me}).
Using standard methods\,\cite{carmichael:book}, we find that the first order coherence function, $G^{(1)}=\langle a^{\dagger}(t)a(t+\tau)\rangle$, decays at a rate $\simeq 2\gamma_{\varphi}$ (details are provided in Appendix \ref{sec:firstordercoherence}). This implies that the linewidth of the resonator spectrum will be a factor of 4 larger than for the resonance at $\omega_{J}\sim\omega_0$ where a single Cooper-pair produces a photon\,\cite{gramich2013}; this is important because it means a signature of the double Cooper-pair tunneling can be found by measuring just the resonator spectrum. The corresponding steady-state occupation number of the resonator
\begin{equation}
\langle \hat{n}\rangle=\left(\frac{\tilde{E}^2_J\Delta_0^3}{3\hbar^2\omega_{J}}\right)^2\frac{1+4\gamma_{\varphi}/\gamma}{\left(\gamma/2\right)^2\left(1+4\gamma_\varphi/\gamma\right)^2+(\delta')^2},\label{eq:h0n}
\end{equation}
grows as $E_J^4$ (to lowest order).
In contrast to the spectral linewidth, the occupation number is only very weakly dependent on low-frequency voltage fluctuations (since $\gamma_{\varphi}/\gamma\ll1$ in typical experimental set-ups\,\cite{hofheinz,gramich2013,wang2017}). In the following we set $\gamma_{\varphi}\rightarrow 0$ for simplicity.

\begin{figure}[t]
\centering
{\includegraphics[width=8.5cm]{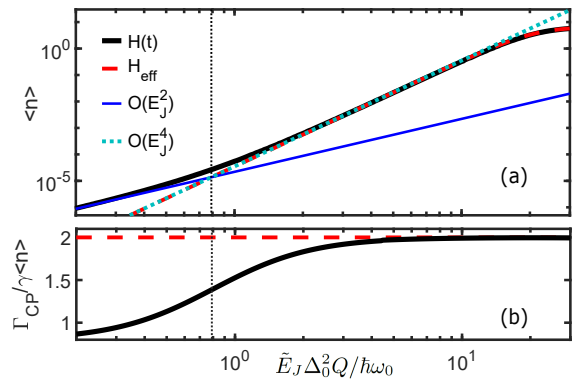}
}
\caption{(a) Time-averaged steady-state photon occupation number using the full Hamiltonian $H(t)$, \eqref{eq:horiginal}, compared with the prediction of $H_{\rm{eff}}$, \eqref{eq:heff}. Also shown are the $\mathcal{O}(E_J^2)$ and $\mathcal{O}(E_J^4)$ contributions to $\langle n\rangle$ obtained from a semiclassical analysis, with a dashed vertical line at $\tilde{E}_J\Delta_0^2Q/\hbar\omega_0=\sqrt{5/8}$ indicating the crossover. (b) Onset of double Cooper-pair transport seen through the ratio  ${\Gamma_{CP}}/{\gamma\langle \hat{n}\rangle}$. The  time-averaged calculation using \eqref{eq:ioriginal} and $H(t)$ (black line) increases slowly and then saturates whereas \eqref{eq:idcteff} predicts a value of precisely 2 (red dashes). The parameter values are: $\omega_J=\omega_0/2$, $Q=1500$, $\Delta_0=0.15$ and $\gamma_{\varphi}=0$.}
\label{fig:two}
\end{figure}

 Comparisons with numerical calculations\,\cite{qutip} using the full Hamiltonian [Eq.\ \ref{eq:horiginal}] in Fig.\ \ref{fig:two}a show that $\langle n\rangle$ does indeed scale as $E_J^4$, but only for intermediate values. For larger $E_J$ values, the contributions at higher order in $E_J$ which are described by \eqref{eq:heff} are required and the photon-number dependent nonlinearities lead to a saturation in photon numbers. However, the behavior at very low $E_J$ is not captured by the effective Hamiltonian \eqref{eq:heff}. This is inevitable because the system is bound to have a period $2\pi/\omega_J$ matching that of the underlying Hamiltonian \eqref{eq:H0} in the limit of very weak $E_J$, whereas the effective Hamiltonian only describes oscillations at $2\omega_J\simeq\omega_0$.

 The low-$E_J$ behavior can be obtained through a physically transparent semiclassical analysis (details of which are  given in Appendix \ref{sec:semiclassical}), utilising the fact that for $\gamma_{\varphi}\rightarrow 0$ the system can be mapped onto a nonlinearly driven oscillator\,\cite{armour2013,meister,armour2017}. This reveals that there is a competition between oscillations with periods $2\pi/\omega_J$ and $\pi/\omega_J$ leading to contributions to (time-averaged) $\langle n\rangle$ that grow as $\tilde{E}_J^2\Delta_0^2$ and $\tilde{E}_J^4\Delta_0^6$ respectively. The contributions have the same weight when $\tilde{E}_J\Delta_0^2Q/\hbar\omega_0=\sqrt{5/8}$ [see Fig.\ \ref{fig:two}a]. 

The instantaneous expectation value of the current flowing through the junction is given by $\langle \hat{I}_{CP}\rangle=2e\langle \dot{N}\rangle$. Since the dissipative terms in the master equation transfer no charge, the current operator is defined by the operator\,\cite{gramich2013,armour2017}
\begin{equation}
{\hat{I}_{CP}(t)}=(2eE_J/\hbar)\sin[\omega_Jt-\varphi+\Delta_0(\hat{a}+\hat{a}^{\dagger})]. \label{eq:ioriginal}
\end{equation}
The expectation value of the current is not stationary, but averaging over a time $T\gg1/\omega_J$ leads to a corresponding expression for the average, or dc, current:
\begin{equation}
\overline{I}_{CP}=\frac{1}{T}\int^{t_0+T}_{t_0}{\rm{d}}t\hat{I}_{CP}(t). \label{eq:idc}
\end{equation}

We can also use the effective Hamiltonian to write down an expression for a time-averaged current operator directly,
\begin{equation}
\overline{I}_{CP}=i\frac{2e}{\hbar}[H_{\rm{eff}},N]\label{eq:idcteff}.
\end{equation}
In terms of the Cooper-pair tunneling rate, $\Gamma_{CP}=\langle\overline{I}_{CP}\rangle/2e$, this expression taken together with \eqref{eq:me} leads to a straightforward relationship in the steady-state: $\Gamma_{CP}/\gamma\langle \hat{n}\rangle=2$. The ratio has this simple integer value because the effective Hamiltonian describes a resonator oscillating at a single frequency (it is stationary in the frame rotating at $2\omega_J$) in which individual photons are always generated (or destroyed) in association with the tunneling of {\emph{two}} Cooper-pairs [see \eqref{eq:heff}]. As Fig.\ \ref{fig:two}b shows, when $\Gamma_{CP}/\gamma\langle \hat{n}\rangle$ is calculated using \eqref{eq:ioriginal} there is excellent agreement with the prediction of the effective Hamiltonian approach at sufficiently large ${E}_J$ values, but it drops below 2 when ${E}_J$ is very small and oscillations at the Josephson frequency can no longer be neglected. In this regime the charge transport is a mixture of processes involving either two or one Cooper-pair(s).

\section{From incoherent to coherent double Cooper-pair tunneling}
To gain an understanding of {\emph{how}} the charge transport takes place, we define a time-averaged current noise\,\cite{bb} for the system through the relation
\begin{eqnarray}
S_{CP}&=&2{\rm{Re}}\int_0^{\infty}{\rm{d}}\tau\int_{t_0}^{t_0+T} \frac{{\rm{d}}t}{T}\left[\langle{\hat{I}_{CP}}(t+\tau){\hat{I}_{CP}}(t)\rangle\right. \nonumber\\
&&\left.-\langle{\hat{I}_{CP}}(t+\tau)\rangle\langle{\hat{I}_{CP}}(t)\rangle\right]. \label{eq:scp}
\end{eqnarray}
When the effective Hamiltonian holds, an equivalent expression for the current noise can be written in terms of the time-averaged current operator, \eqref{eq:idcteff}.
The corresponding Fano factor, $F_{CP}=S_{CP}/(2e\langle \overline{I}_{CP}\rangle)$, compares the noise to that of a Poissonian process involving a single Cooper-pair\,\cite{armour2017}, providing a convenient way of characterising the behavior.

\begin{figure}[t]
\centering
{
\includegraphics[width=8.0cm]{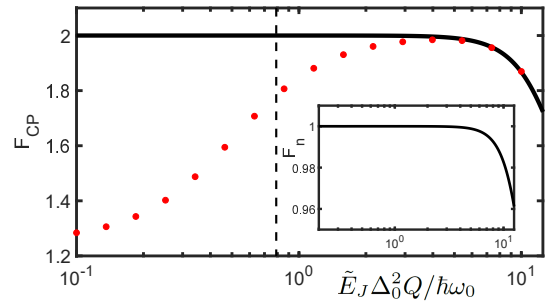}
}
\caption{Variation of $F_{CP}$ with $E_J$. Numerical integrations using the time-dependent Hamiltonian [Eq.\ \eqref{eq:horiginal}] (points) are compared with calculations using the effective Hamiltonian [Eq.\ \eqref{eq:heff}] (line); a dashed vertical line indicates $\tilde{E}_J\Delta_0^2Q/\hbar\omega_0=\sqrt{5/8}$. The inset shows the corresponding behavior of the fluctuations in the resonator occupation number, $F_n$, calculated using \eqref{eq:heff}. Parameters match those used in Fig.\ \ref{fig:two}.}
\label{fig:three}
\end{figure}

As Fig.\ \ref{fig:three} shows, using the effective Hamiltonian leads to a value of $F_{CP}$ which tends to 2 in the limit of small $E_J$; this signifies incoherent tunneling of {\emph{two}} Cooper pairs\,\cite{clerk2} (i.e.\ charge $4e$).  For larger $E_J$ values, $F_{CP}$ drops. We know that in this regime on average two Cooper-pairs tunnel for each photon entering the resonator so this implies that the transport of {\emph{pairs}} of Cooper-pairs becomes coherent\,\cite{grabertepl}. This is accompanied by sub-Poissonian photon statistics within the resonator (i.e.\ $F_n=(\langle\hat{n}^2\rangle-\langle\hat{n}\rangle^2)/\langle \hat{n}\rangle<1$, see the inset of Fig.\ \ref{fig:three}), and is similar to the transition from incoherent to coherent tunneling of {\emph{single}} Cooper pairs\,\cite{gramich2013,armour2017} that occurs for $\omega_J\simeq\omega_0$. For low values of $E_J$ the effective Hamiltonian approach fails and numerical calculations using \eqref{eq:scp} and the full time-dependent Hamiltonian show that $F_{CP}$ drops below 2, but in this case it is because single Cooper-pair tunneling processes have become important.

The regime where charge transport is almost entirely due to incoherent double Cooper-pair tunneling (IDCPT), and hence $F_{CP}\sim 2$, maps onto the domain of validity of \eqref{eq:H0}: set below by the crossover to  (off-resonant) single-Cooper pair tunneling events ($\tilde{E}_J\Delta_0^2Q/\hbar\omega_0\sim \sqrt{5/8}$) and above by the onset of strong effective nonlinearities ($4\Delta_0^2\langle \hat{n}\rangle\sim 1$). Hence, we expect IDCPT to dominate when
$\sqrt{5/8}\ll \tilde{E}_J\Delta_0^2Q/\hbar\omega_0\ll\sqrt{3Q/8}$, which means that it will only be well-separated from other transport regimes for weak damping, $Q\gg 1$.

\section{Single photon nonlinearity and photon blockade}

We now turn to the strongly non-classical behavior of the system which emerges when $\Delta_0\sim 1$. Of particular interest is the behavior of the matrix element $\langle 1|H_{{\rm{eff}}}|2\rangle$, for which a closed form expression can be derived analytically (see Appendix \ref{sec:effham}). If this is zero the system becomes trapped within the two-state basis of the $n=0,1$ Fock states\,\cite{gramich2013,clerk,simon,esteveth}. As a consequence, the corresponding correlation function $g^{(2)}(0)=\langle \hat{a}^{\dagger}\hat{a}^{\dagger}\hat{a}\hat{a}\rangle/\langle \hat{a}^{\dagger}\hat{a}\rangle^2$ vanishes indicating photon-blockade and the system can function as a single photon source\,\cite{antibunched}.

\begin{figure}[t]
\centering
{
\includegraphics[width=8.0cm]{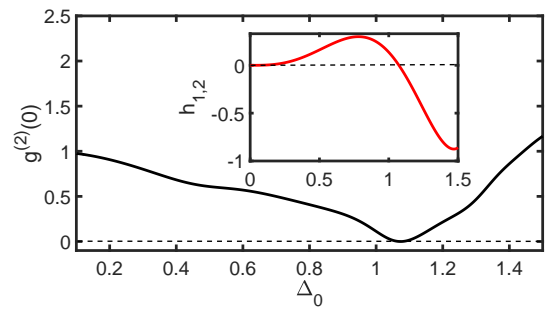}
}
\caption{Variation of $g^{(2)}(0)$ [calculated using \eqref{eq:heff}] and $h_{1,2}=(4\hbar\omega_J/\tilde{E}_J^2)|\langle 1|H_{{\rm{eff}}}|2\rangle|$ with $\Delta_0$. 
In the main plot, 
$\tilde{E}_J/\hbar\omega_0=0.1$, $\omega_J=\omega_0/2$, $Q=500$ and $\gamma_{\varphi}=0$.}
\label{fig:four}
\end{figure}

Despite their apparent complexity the matrix elements of the effective Hamiltonian \eqref{eq:heff} do have zeros, implying destructive interference of the many processes which contribute (Fig.\ \ref{fig:cpoint}c), and hence there is a strong photon blockade effect as Fig.\ \ref{fig:four} illustrates. The zero of $\langle 1|H_{{\rm{eff}}}|2\rangle$ with lowest $\Delta_0$ occurs at $\Delta_0\simeq 1.07$. Surprisingly, this is a significantly lower value than the corresponding matrix-element zero that occurs for the single photon resonance where Cooper-pairs tunnel individually ($\omega_J\simeq\omega_0$) which occurs at $\Delta_0=\sqrt{2}$\,\cite{gramich2013,antibunched}.

\section{Conclusions}
In conclusion, we have analysed charge transport and photon emission in a JJ-resonator system biased so that the Josephson frequency is just half the resonator frequency. As the Josephson energy is increased, the underlying dynamics of the system crosses over from oscillations at the Josephson frequency to the resonator frequency, accompanied by a corresponding crossover in charge transport from single to double Cooper-pair tunneling. By deriving an effective Hamiltonian description, we found that whilst double Cooper-pair transport is incoherent when it first begins to dominate, a regime of coherent double-Cooper pair tunneling emerges at larger Josephson energies. For large resonator impedances the system displays a photon blockade which could be exploited as a single photon source.

Whilst double Cooper-pair processes are higher-order in both the Josephson energy and resonator impedance than their single Cooper-pair counterparts,  the values of both  these quantities can be tuned in experiments within broad ranges (e.g.\ $\Delta_0$ up to $\sim 1$\,\cite{antibunched} and $E_J$ beyond $\hbar\omega_0$\,\cite{chen2014}), making the novel charge transport and photon-emission regimes we describe readily accessible with current device architectures\,\cite{chen2014,cassidy,westig17,amplifier,singlephotonsource,antibunched}. Our work opens the way for future work exploring how charge transport might be controlled via microwave cavities and could also stimulate interest in overbias emission in superconducting STM\,\cite{SCSTM}, a potentially very fruitful direction given the rich behavior seen in normal state STM\,\cite{OBE2,OBE3,OBE4,OBE5}.

{\emph{Note Added}}: In the final stages of preparing our paper another study appeared which also involves double Cooper-pair tunneling\,\cite{devoret}, albeit in a very different context.

\section*{Acknowledgements} We thank J. Ankerhold, B. Kubala, B. Lang and F. Portier for useful discussions.
    This work was supported by the Engineering and Physical Sciences Research Council [grant number  EP/P510592/1] through a studentship held by WTM, through a  Leverhulme Trust Research Project Grant (RPG-2018-213), by the German Excellence
    Initiative through the Zukunftskolleg and the Deutsche
    Forschungsgemeinschaft (DFG) through the SFB 767.

\appendix

\begin{widetext}
\section{Effective Hamiltonian}
\label{sec:effham}

The operators $\hat{\mathcal{G}}(\hat{n},\Delta_0)\hat{n}$ and $\hat{a}\hat{\mathcal{F}}(\hat{n},\Delta)$ that appear in the expression for the effective Hamiltonian [Eq.\ \eqref{eq:heff}] are defined by the relations,
\begin{eqnarray}
\hat{\mathcal{G}}(\Delta_0,\hat{n})\hat{n}&=&\sum_{p=1}^{+\infty}\frac{4p}{4p^2-1}[\hat{A}_p,\hat{A}^{\dagger}_p]\\
\hat{a}\hat{\mathcal{F}}(\Delta_0,\hat{n})&=&\sum_{p=0}^{+\infty}\frac{(-1)^p}{2p+1}[\hat{A}_p,\hat{A}^{\dagger}_{p+1}]
\end{eqnarray}
with $\hat{A}_p=(\hat{a}^{\dagger})^p\hat{K}_p$,
where the Hermitian operator $\hat{K}_p$ is a function of the number operator only and is defined as
\begin{equation}
\hat{K}_p=:\frac{J_p(2\Delta_0\sqrt{\hat{n}})}{\hat{n}^{p/2}}:=\sum_{m=0}^{+\infty}\frac{(-1)^m\Delta_0^{2m+p}(\hat{a}^{\dagger})^m\hat{a}^m}{m! (m+p)!}.
\end{equation}

The effective Hamiltonian [Eq.\ \eqref{eq:heff}] can also be expressed directly in terms of the Fock state basis,
\begin{equation}
H_{\rm{eff}}=\sum_{q=0}^{+\infty}(q\hbar \delta+\delta E_q)|q\rangle\langle q|+i\sum_{q=0}^{+\infty}\left[M_{q,q+1}|q\rangle\langle q+1|{\rm{e}}^{-2i\varphi}-{\rm{h.c.}}\right],
\end{equation}
with the matrix elements defined as
\begin{eqnarray}
\delta E_{q}&=&\frac{\tilde{E}_J^2}{4\hbar\omega_J}\left\{\sum_{p=1}^q\frac{4p}{4p^2-1}\left[\frac{\kappa^2_p(q-p)q!}{(q-p)!}\right]-\sum_{p=1}^{+\infty}\frac{4p}{4p^2-1}\left[\frac{\kappa^2_p(q)(q+p)!}{q!}\right]\right\}\\
M_{q,q+1}&=&\frac{\tilde{E}_J^2}{4\hbar\omega_J}\left\{\sum_{p=0}^q\frac{(-1)^p}{2p+1}\frac{\sqrt{q!(q+1)!}}{(q-p)!}\kappa_p(q-p)\kappa_{p+1}(q-p)-\sum_{p=0}^{+\infty}\frac{(-1)^p}{2p+1}\frac{(q+p+1)!}{\sqrt{q!(q+1)!}}\kappa_{p+1}(q)\kappa_{p}(q+1)\right\},
\end{eqnarray}
where $\kappa_p(q)$ is the $q$-th eigenvalue of $\hat{K}_p$ (i.e.\ $\hat{K}_p|q\rangle=\kappa_p(q)|q\rangle$), which is given by
\begin{equation}
\kappa_p(q)=q!\sum_{n=0}^{q}\frac{(-1)^n\Delta_0^{2n+p}}{n!(n+p)!(q-n)!}.
\end{equation}
Using this matrix representation, one can then derive closed form expressions for specific matrix elements of the Hamiltonian. In particular, we find
\begin{equation}
M_{1,2}=\frac{\tilde{E}_J^2}{4\sqrt{2}\hbar\omega_J}\left[\Delta_0{\rm{e}}^{-\Delta_0^2}\left(\frac{2}{3}\Delta_0^4-\frac{10}{3}\Delta_0^2+\frac{3}{2}\right)+\sqrt{\pi}{\rm{erf}}(\Delta_0)\left(\frac{2}{3}\Delta_0^6-3\Delta_0^4+\frac{7}{2}\Delta_0^2-\frac{3}{4}\right)\right]
\end{equation}
with ${\rm{erf}}(x)$ the (Gauss) error function.  As discussed in the main text, this has its first zero at $\Delta_0\sim 1.07$ [see the inset of Fig.\ \ref{fig:four}].

\section{First-order coherence function}
\label{sec:firstordercoherence}

In this section we outline the calculation of the first-order coherence function, a similar calculation for the single-Cooper pair resonance (where $\omega_J\simeq \omega_0$) is discussed in Ref.\ \onlinecite{gramich2013}. Starting from Eqs.\ \eqref{eq:me} and \eqref{eq:H0}, we obtain the equations of motion
\begin{eqnarray}
\frac{d}{dt}\langle{\hat{a}}\rangle&=&-(i\delta'+\gamma/2)\langle \hat{a}\rangle-X\langle {\rm{e}}^{2i\varphi}\rangle\\
\frac{d}{dt}\langle {\rm{e}}^{2i\varphi}\rangle&=&-2\gamma_{\varphi}\langle {\rm{e}}^{2i\varphi}\rangle.
\end{eqnarray}
where $X=\tilde{E}^2_J\Delta_0^3/(3\hbar^2\omega_J)$. Hence, using the regression formula\,\cite{carmichael:book} we find
\begin{eqnarray}
\langle \hat{a}^{\dagger}(t){\rm {e}}^{2i\varphi}(t+\tau)\rangle&=&\langle \hat{a}^{\dagger}(t){\rm {e}}^{2i\varphi}(t)\rangle{\rm{e}}^{-2\gamma_{\varphi}\tau},\\
\langle \hat{a}^{\dagger}(t)a(t+\tau)\rangle&=&\langle \hat{a}^{\dagger}(t)\hat{a}(t)\rangle {\rm{e}}^{-(\gamma/2+i\delta')\tau}-X\frac{\langle \hat{a}^{\dagger}(t){\rm {e}}^{2i\varphi}(t)\rangle}{\gamma/2-2\gamma_{\varphi}+i\delta'}\left({\rm{e}}^{-2\gamma_{\varphi}\tau}-{\rm{e}}^{-(\gamma/2+i\delta')\tau}\right).
\end{eqnarray}
Using the steady-state values ($t\rightarrow\infty$)
\begin{eqnarray}
\langle \hat{n}\rangle&=&\langle \hat{a}^{\dagger}\hat{a}\rangle=\frac{X^2(1+4\gamma_{\varphi}/\gamma)}{(\gamma/2+2\gamma_{\varphi})^2+(\delta')^2}\\
\langle \hat{a}^{\dagger}{\rm {e}}^{2i\varphi}\rangle&=&\frac{-X}{\gamma/2+2\gamma_{\varphi}-i\delta'},
\end{eqnarray}
leads to
\begin{eqnarray}
\langle \hat{a}^{\dagger}\hat{a}(\tau)\rangle&=&\langle \hat{n}\rangle{\rm{e}}^{-(i\delta'+\gamma/2)\tau}+\frac{X^2\left[{\rm{e}}^{-2\gamma_{\varphi}\tau}-{\rm{e}}^{-(i\delta'+\gamma/2)\tau}\right]}{\left[i\delta'+(\gamma/2-2\gamma_{\varphi})\right]\left[-i\delta'+(\gamma/2+2\gamma_{\varphi})\right]}.
\end{eqnarray}
Finally, assuming $\gamma_{\varphi}/\gamma\ll 1$, we can simplify this to
\begin{equation}
\langle \hat{a}^{\dagger}\hat{a}(\tau)\rangle\simeq\langle \hat{n}\rangle{\rm{e}}^{-2\gamma_{\varphi}\tau}.
\end{equation}

\section{Semiclassical Analysis}
\label{sec:semiclassical}

A simple semiclassical model for the system (similar in spirit to that discussed in Ref.\ \onlinecite{meister}) is obtained from Eqs.\ \eqref{eq:horiginal} and \eqref{eq:me} by making the ansatz that the resonator is in a coherent state $|\alpha\rangle$. Taking the limit $\gamma_{\varphi}\rightarrow 0$ and setting $\varphi=0$ for convenience, we find
\begin{equation}
\dot{\alpha}=-\left(i\omega_0+\frac{\gamma}{2}\right)\alpha-\frac{i\tilde{E}_J\Delta_0}{\hbar}\sin[\omega_Jt+\Delta_0(\alpha+\alpha^*)]. \label{eq:alpha}
\end{equation}
For $\omega_J\simeq \omega_0/2$ and the very smallest $E_J$ values the system will behave like a linear oscillator subject to two off-resonant drives so that in the limit of long times, $\alpha\simeq\alpha_-{\rm{e}}^{-i\omega_Jt}+\alpha_+{\rm{e}}^{+i\omega_Jt}$ with constants $\alpha_{\pm}$. However, for slightly larger $E_J$ values the nonlinearity will up-convert the oscillations (with amplitudes $\alpha_{\pm}$) at frequency $\omega_J$ into an effective drive near the resonant frequency ($\omega_0\simeq 2\omega_J$), to take these into account we assume a solution of the form $\alpha=\alpha_0{\rm{e}}^{-2i\omega_Jt}+\alpha_-{\rm{e}}^{-i\omega_Jt}+\alpha_+{\rm{e}}^{i\omega_Jt}$, substituting this into \eqref{eq:alpha} and assuming harmonic balance leads to the relations
\begin{eqnarray}
\alpha_0&=&-i\frac{\tilde{E}_J\Delta_0^2}{2\hbar}\frac{(\alpha_-+\alpha_+^*)}{i(\omega_0-2\omega_J)+\gamma/2}\label{eq:upconvert}\\
\alpha_-&=&\frac{\tilde{E}_J\Delta_0}{2\hbar}\frac{1-i\Delta_0\alpha_0}{i(\omega_0-\omega_J)+{\gamma/2}},\\
\alpha_+&=&-\frac{\tilde{E}_J\Delta_0}{2\hbar}\frac{1+i\Delta_0\alpha_0^*}{i(\omega_0+\omega_J)+{\gamma/2}}.
\end{eqnarray}
Using the fact that $\omega_J\simeq \omega_0/2$, assuming $\gamma/2\ll\omega_J$ and working to 4th order in $\tilde{E}_J$ (and 6th order in $\Delta_0$) leads to the approximate expression
\begin{eqnarray}
\overline{\langle n\rangle}&\simeq& {\overline{|\alpha|}}^2=|\alpha_0|^2+|\alpha_+|^2+|\alpha_-|^2 \\
&\simeq&\frac{\tilde{E}^2_{J}\Delta_0^2}{2\hbar^2}\left[\frac{\omega_0^2+\omega_J^2}{(\omega_0^2-\omega_J^2)^2}\right]\left\{1+\frac{\tilde{E}^2_{J}\Delta_0^4}{2\hbar^2}\frac{\omega_0^2}{[\omega_0^2+\omega_J^2][(\omega_0-2\omega_J)^2+\gamma^2/4]}\right\} \label{eq:upconvert2}
\end{eqnarray}
where the bar implies a time average. The oscillations at $\pm\omega_J$ give rise to a contribution to $\overline{\langle n\rangle}$ that grows as $\sim {\tilde{{E}}}_J^2\Delta_0^2$, whilst the oscillations at frequency $2\omega_J$ give rise to one that grows as $\sim {\tilde{{E}}}_J^4\Delta_0^6$. The crossover between these two components is obtained by equating the two terms in the braces.
As \eqref{eq:upconvert} and \eqref{eq:upconvert2} make clear, the amplitude oscillating at $2\omega_J$ is indeed an upconversion of the oscillations at $\omega_J$. Furthermore, one cannot neglect the most off-resonant component ($\alpha_+$), doing so leads to a value for $\overline{|\alpha|}^2$ which is very noticeably less accurate. Notice that the ${\mathcal{O}}(E_J^4)$ component of the average occupation number,
\begin{equation}
|\alpha_0|^2\simeq \left(\frac{\tilde{E}^2_{J}\Delta_0^3}{3\hbar^2\omega_J}\right)^2\frac{1}{\delta^2+\gamma^2/4}, \label{eq:o4}
\end{equation}
matches Eq.\ \eqref{eq:h0n}  in the limit $\gamma_{\varphi}\rightarrow 0$, up to higher-order corrections (in $\Delta_0$ and $E_J$) arising from the frequency shift $\delta'-\delta$. The frequency shift can also be obtained within the semiclassical approach, but through a calculation that goes to higher order\,\cite{meister}. In Fig.\ \ref{fig:two}, Eq.\ \ref{eq:o4} is the $\mathcal{O}(E_J^4)$ expression plotted, whilst the  $\mathcal{O}(E_J^2)$ one is the corresponding part of \eqref{eq:upconvert2},
\begin{equation}
|\alpha_+|^2+|\alpha_-|^2 \simeq\frac{\tilde{E}^2_{J}\Delta_0^2}{2\hbar^2}\left[\frac{\omega_0^2+\omega_J^2}{(\omega_0^2-\omega_J^2)^2}\right].
\end{equation}

In the semiclassical description, the time-averaged current is given by
\begin{equation}
\frac{\overline{I}_{CP}}{2e}=\frac{\tilde{E}_J}{\hbar}\overline{\sin[\omega_Jt+\Delta_0(\alpha+\alpha^*)]}.
\end{equation}
Using the same ansatz for $\alpha$ as above,  together with the assumptions $\omega_J\simeq \omega_0/2$, $\gamma/2\ll\omega_J$ and again working to 4th order in $\tilde{E}_J$ (and 6th order in $\Delta_0$)  leads to
\begin{eqnarray}
\frac{\overline{I}_{CP}}{2e}&\simeq&\frac{\gamma\tilde{E}_J^2\Delta_0^2\omega_0\omega_J}{\hbar^2(\omega_0^2-\omega_J^2)^2}
+\frac{\gamma\tilde{E}^4_{J}\Delta_0^6}{2\hbar^4}\frac{\omega_0^2}{(\omega_0^2-\omega_J^2)^2[(\omega_0-2\omega_J)^2+\gamma^2/4]}. \label{eq:sccurr}
\end{eqnarray}
Hence in the limit $E_J\rightarrow 0$ the time-averaged current to photon number ratio will be
\begin{equation}
\frac{\langle \overline{I}_{CP}\rangle}{2e\gamma\langle n\rangle}=\frac{2\omega_0\omega_J}{\omega_0^2+\omega_J^2}.
\end{equation}
This matches the drop below unity seen in the upper panel of Fig.\ \ref{fig:two}. On the other hand, if just the $\tilde{E}_J^4$ contributions are included then the ratio is simply 2.

\end{widetext}

\end{document}